\documentclass[12pt]{article}
 \pdfoutput=1
\usepackage{graphicx,wrapfig}

\usepackage[english]{babel}
\usepackage{fancyhdr}
\usepackage{amsmath}
\usepackage{amssymb}
\usepackage{amsfonts}
\usepackage{psfrag}
\usepackage[applemac]{inputenc}

\usepackage{enumerate}

\newcommand{\be}{\begin{equation}}
\newcommand{\ee}{\end{equation}}
\newcommand{\bea}{\begin{eqnarray}}
\newcommand{\eea}{\end{eqnarray}}
\newcommand{\bag}{\begin{align}}
\newcommand{\eag}{\end{align}}

\newcommand{\GeV}{\,\mathrm{GeV}}
\newcommand{\TeV}{\,\mathrm{TeV}}

\newcommand{\eq}[1]{Eq.~(\ref{#1})}

\newcommand{\pslash}{\!\not\! p}

\begin{document}
\baselineskip=18pt
\setcounter{footnote}{0}
\setcounter{figure}{0}
\setcounter{table}{0}


\begin{titlepage}
\begin{flushright}
May 2012
\end{flushright}
\vspace{.3in}

\begin{center}
\vspace{1cm}

{\Large \bf
The
  Composite Higgs and Light Resonance Connection}

\vspace{1.2cm}

{\large  Alex Pomarol\,$^a$ and Francesco Riva\,$^b$\\}
\vspace{.8cm}
{\it {$^a$\, Departament  de F\'isica, Universitat Aut{\`o}noma de Barcelona, 08193~Bellaterra,~Barcelona}}\\
{\it {$^b$\, IFAE, Universitat Aut{\`o}noma de Barcelona, 08193~Bellaterra,~Barcelona}}\\

\vspace{.4cm}

\end{center}
\vspace{.8cm}

\begin{abstract}
\medskip
\noindent
Weinberg sum-rules  have  been used in the past  to successfully predict  the electromagnetic  contribution to  the charged-pion mass as a function of  the meson masses. Following   the same approach
we calculate in  the minimal composite Higgs model (MCHM)
 the Higgs mass as a function of the  fermionic resonance masses.
The simplicity of the method allows us to study several versions of the MCHM and show
that  a Higgs  with a mass around 125 GeV requires, quite  generically,
 fermionic resonances below the TeV, and therefore accessible at the  LHC. 
We also examine the couplings  of the Higgs to the SM fermions  and calculate their 
deviation from the SM value.
\end{abstract}

\bigskip

\end{titlepage}

\section{Introduction}
LHC Higgs boson searches \cite{ATLAS:2012ae,Chatrchyan:2012tx}
 have cornered  the Higgs into small  mass windows. Combining these results with LEP \cite{Barate:2003sz},
one can conclude, at $95\%$ CL, that  the Higgs mass must lie in the narrow region  $114.4\ \text{GeV}< m_h<127\ \text{GeV}$, or else it  must be heavier than $\sim 600$ GeV. More interestingly,  the recently observed excess of events points towards a Higgs with   $m_h\sim 125$ GeV 
and couplings (relatively) close
to the SM values.

The existence of  a light Higgs  has important implications on models for the electroweak scale.
The most interesting extensions of the SM in which  a light Higgs is predicted are 
 composite Higgs 
and supersymmetric models.
In composite Higgs models  the Higgs, similarly to pions in QCD,
is light since it is  a pseudo-Goldstone boson (PGB) arising from a new strong sector at TeV energies 
\cite{GK,Agashe:2004rs}. Its mass is generated at the loop level through interactions with the SM states, which do not respect the (non-linearly realized) global symmetry which protects the Higgs mass. Since these loops are sensitive to the resonances of the strong sector, in these scenarios the Higgs mass is related to the  mass of the heavy resonances.
The exact relation is however difficult to calculate due to the intractable strong dynamics.
One possibility  to quantify this relation is to use  AdS/CFT techniques
\cite{Agashe:2004rs,Contino:2006qr}.
This has led to precise relations  between the Higgs mass and 
the resonance masses, predicting  that,   in certain models,  a  light Higgs  necessary 
requires   fermionic resonances   below the TeV \cite{Contino:2006qr}.
Another possibility, recently explored, is to use  deconstructed versions of the model, leading to
similar conclusions \cite{Panico:2011pw,DeCurtis:2011yx}.

The purpose of this article is to analyze the relation between the Higgs mass and the fermionic resonance masses making  use of  the Weinberg sum-rules  \cite{Weinberg:1967kj} which
 provide high-energy constraints on correlators  (or spectral functions).
In QCD, the Weinberg sum-rules, when combined with the assumption that only the lowest-mass resonances dominate the correlators,   have led to important relations between  physical quantities.
One celebrated example is the electromagnetic contribution to the charged-pion mass
\cite{Das:1967it}, 
where a  successful prediction was obtained,
$m^2_{\pi^+}-m^2_{\pi^0}\simeq 3\alpha m_\rho^2\log2/(2\pi)$, relating the  PGB mass
to  the resonance masses.
Following a similar approach we will calculate the mass of a PGB Higgs.
We will focus on  the  Minimal Composite Higgs model (MCHM) \cite{Agashe:2004rs}, 
generically defined as the  model where the Higgs arises as a PGB from a 
strong sector whose   global symmetry   SO(5) is broken dynamically down to SO(4).
Imposing the equivalent of the QCD Weinberg sum-rules in the large-$N_c$ limit,  and assuming 
that the correlators are saturated by the lightest resonances,
we will  find simple expressions relating the Higgs mass to the fermionic resonance masses.
The simplicity of the method will  allow us to go beyond the minimal versions of the MCHM
and study other scenarios.
With the exception of very particular cases, we will show that a light Higgs necessarily implies fermionic resonances
below the TeV, accessible to the LHC.

In section \ref{sec:HIGGSMASS} we  explain   
our  procedure and show how to   calculate  the Higgs mass for the MCHM$_5$ and MCHM$_{10}$.
In section \ref{sec:HiggsCouplings} we   extend this calculation to other MCHM  and derive
a generic lower-bound on the Higgs mass.
 In section \ref{sec:Conclusions} we  summarize our results.
In  Appendix A  we give the explicit relations between the 
top-quark form-factors and the correlators of the strong sector, while 
in Appendix B we give the effective  lagrangian of the top in  certain MCHM models of interest.

\vskip1cm
{\bf Note added:} While this work was in preparation, Ref.~\cite{Marzocca:2012zn} appeared, where the Weinberg sum-rules are also used to link the Higgs and fermion resonance masses and  some of the formulas presented here are also derived.

\section{The Higgs mass   in the MCHM}
\label{sec:HIGGSMASS}

In this section, we want to calculate the Higgs mass as a function of the  resonance masses of the strong sector
in different realizations of the MCHM. We will work in the unitary gauge where only the physical Higgs $h$ is kept and the SM Goldstones
are gauged away. 
We start with the calculation of the  gauge contribution  to the Higgs potential, that  follows  closely     the  original  calculation of the electromagnetic contribution to the charged-pion mass \cite{Das:1967it}. 
 Then we compute the fermion contribution which, due to the large top-quark Yukawa coupling, is typically dominant.

\subsection{Gauge contributions to the Higgs potential}

Working in the limit $g'\rightarrow0$,
the  SM gauge contribution   
arising from loops of SU(2)$_L$ gauge bosons is given by~\cite{Agashe:2004rs}
\begin{equation}
V_{gauge}(h) = \frac{9}{2} \int\! \frac{d^4p}{(2\pi)^4} \log\left(\Pi_0(p) +\frac{s^2_h}{4}\, \Pi_1(p) \right)\, ,
\label{pot0}
\end{equation}
where  $s_h\equiv \sin h/f$, being $f$ the PGB decay-constant, and 
$p$ the Euclidian 4-momentum.
We also have 
\begin{equation}
\Pi_0(p)=\frac{p^2}{g^2}+\Pi_a(p)\ , \qquad
\Pi_1(p)=2\big[\Pi_{\hat a}(p)-\Pi_a(p)\big]\, ,
\label{ff}
\end{equation}
where  $g$ is the gauge coupling and $\Pi_{a}(p)$ is the two-point function of the SO(4) conserved current in momentum space,
  $\Pi_{a}\sim\langle J_a J_a \rangle$, and similarly $\Pi_{\hat a}$ for the current associated to the broken generators in SO(5)/SO(4);
for the precise definitions see Ref.~\cite{Agashe:2004rs}.
In a large-$N$ expansion, that we will assume here, these form factors can be written as an infinite  sum over narrow resonances:
\begin{equation}
\Pi_{a}(p)=  p^2
 \sum_n\frac{F^2_{\rho_n}}{p^2+m^2_{\rho_n}}\, ,\qquad
\Pi_{\hat a}(p)= p^2
 \sum_n\frac{F^2_{a_n}}{p^2+m^2_{a_n}} + \frac{1}{2} f^2\, ,
\label{valargen}
\end{equation}
where $\rho_n$ and $a_n$
are vector resonances  coming  respectively in   $\bf 6$-plets  and $\bf 4$-plets of SO(4),
and      $F_{\rho_n,a_n}$ are  referred to as the decay-constants of these resonances.

The  Higgs-dependent part of the  potential  \eq{pot0} is  expected to be 
finite.
Indeed, according to  the operator product expansion,
  the form factor $\Pi_1(p)$  must drop at large $p$ as  $\sim \langle {\cal O}\rangle/p^{d-2}$, 
  where $\cal O$ is the lowest dimension $d$ operator of the strong sector  responsible for the SO(5)\ $\rightarrow$\ SO(4)  breaking. 
In large-$N_c$  QCD, in the limit of massless  quarks, 
we have $\langle {\cal O}\rangle\sim \langle q\bar q\rangle^2$ and then $d=6$,  with
the left-right correlator $\Pi_{LR}(p)=\Pi_V-\Pi_A\rightarrow \langle q\bar q\rangle^2/p^4$  being the equivalent of our $\Pi_1(p)$.
We assume that in the TeV strong sector $d>4$, meaning that
the integral    $\int d^4p\,   \Pi_1(p)/\Pi_0(p)$  is convergent for  $\Pi_0\sim p^2$,
assuring 
the finiteness of  the Higgs-dependent part of the  potential  \eq{pot0}.
This convergence is equivalent to imposing  a set of requirements on $\Pi_1(p)$,  usually   known  as 
the Weinberg sum-rules \cite{Weinberg:1967kj}. These are
\be
\lim_{p^2\to\infty} \Pi_1(p)= 0\ ,\qquad\qquad 
\lim_{p^2\to\infty} p^2\Pi_1(p)= 0\, ,
\label{sumrules}
\ee
that give two constraints to be fulfilled by  the decay constants  and masses in  \eq{valargen}. 
Following  Ref.~\cite{Das:1967it},  we can now make the extra assumption  of  truncating the infinite sum in  \eq{valargen}
to include only the minimal    number of resonances needed
 to satisfy the  sum-rules \eq{sumrules}.
 One can easily realize that only two are needed,
 $ \rho_1\equiv \rho$ and $a_1$.
Using  the two constraints \eq{sumrules}
we can determine $F_\rho$ and $F_{a_1}$, and then calculate $\Pi_1$ as a function of the two resonance masses~\footnote{This result is straightforward to obtain in the following alternative way. 
Requiring that $\Pi_{1}$ has two poles  corresponding to the two massive
resonances implies that   the denominator of $\Pi_1$ must be $(p^2+m_\rho^2)(p^2+m_{a_1}^2)$; the numerator can easily be obtained by requiring  $\Pi_1(0)=f^2$.}:
\be
\Pi_1(p)=\frac{f^2m_\rho^2m_{a_1}^2}{(p^2+m_\rho^2)(p^2+m_{a_1}^2)}\, .
\label{pi1app}
\ee
\eq{pi1app}
  can now be used to obtain the gauge contribution to the Higgs potential  \eq{pot0}.
In an expansion $g^2\ll 1$, we have
\be
V(h)= \alpha s^2_h+ \beta s^4_h+\cdots\, ,
\ee
where
\bea
\alpha&=&\frac{9g^2f^2m^2_\rho m_{a_1}^2}{128\pi^2(m_{a_1}^2-m^2_\rho)}\log\left(\frac{m_{a_1}^2}{m^2_{\rho}}\right)\, ,\label{gaugecon1}
\\
\beta&=&-\frac{9g^4f^4}{1024\pi^2}\left[
 \log\left(\frac{m_{a_1}m_{\rho}}{m_W^2}\right)
-\frac{(m^4_{a_1}+m^4_{\rho})}{(m_{a_1}^2-m^2_\rho)^2}\right.\nonumber\\
&&\quad\quad\quad\quad\quad
  -\left.\frac{(m^2_{a_1}+m^2_{\rho})(m^4_{a_1}-4m^2_{a_1}m^2_{\rho}+m^4_{\rho})}{2(m_{a_1}^2-m^2_\rho)^3}\log\left(\frac{m_{a_1}^2}{m^2_{\rho}}\right)\right]
\, ,
\label{gaugecon2}
\eea
and in the calculation of  $\beta$ the infrared divergence  has been regularized with  the $W$ mass. 
Notice that, being $\alpha$ positive, the gauge contribution alone cannot induce electroweak symmetry  breaking (EWSB).

\subsection{Top contributions in the MCHM$_5$}

We can now  repeat the same procedure for the fermionic contributions to the Higgs potential, concentrating on the one from the top quark, which
is usually the most important one and generates a Higgs potential  with an EWSB minimum.

As in  Ref.~\cite{Agashe:2004rs}, we  will consider  models in which 
 the SM fermions couple to the strong sector by mixing with  fermionic operators.
These mixings are defined by the embedding
of  the SM fermions into SO(5) spurion fields (see Appendix A).
In this section   we will  work in the MCHM$_5$
  \cite{Contino:2006qr}
where   the left-handed and right-handed top, $t_L$ and $t_R$,  
are respectively    embedded in two  spurions  in the   $\bf r_L=5$ and $\bf r_R=5$ representation of SO(5).
The (non-local)  effective theory for the top quark, at the quadratic order, can be written  in momentum space as
  \begin{equation}
{\cal L}_{\rm eff}=
  \bar t_L \pslash \left(\Pi^{t_L}_0(p)
 +\frac{s_h^2}{2}\,\Pi^{t_L}_1(p)\right)t_L+
  \bar t_R \pslash \left(\Pi^{t_R}_0(p)
 +{c_h^2}\, \Pi^{t_R}_1(p)\right)t_R+
\left(  \frac{s_hc_h}{\sqrt{2}}\,\bar t_L  M^t_1(p)
t_R+h.c.\right)\, ,
 \label{efflag}
\end{equation}
where the form factors $\Pi^{t_{L,R}}_{0,1}(p)$ and $M^t_1(p)$ encode the strong sector dynamics.
The top contribution to the Higgs potential is then   \cite{Contino:2006qr}~\footnote{We are working in a large-$N$ expansion and  neglect   contributions coming from form factors involving four or more top-quarks.}
\begin{equation}
V_{top}(h) =             -2 N_c \int\!\frac{d^4p}{(2\pi)^4}
\log\left[
        - p^2 \left( \Pi_0^{t_L} + \frac{s^2_h}{2}\, \Pi_1^{t_L} \right)
         \left( \Pi_0^{t_R} +{c^2_h}\, \Pi_1^{t_R} \right) 
                    -\frac{s^2_h c^2_h}{2} |M_1^{t}|^2 \right] \, ,
\label{potF}
\end{equation}
where $N_c=3$ and, as   shown in Appendix A, the top-quark form factors can be written as a function of the correlators of the fermionic operators decomposed in SO(4)-representations: 
\bea
&&\Pi_{0}^{t_L}(p)   =1+  \Pi^L_{Q_4}(p)\ ,\qquad\qquad 
     \Pi_{1}^{t_L}(p)   =  \Pi^L_{Q_1}(p)-\Pi^L_{Q_4}(p) \, ,  \nonumber\\
&&\Pi_{0}^{t_R}(p)  =1+  \Pi^R_{Q_4}(p)\ ,\qquad\qquad     
\Pi_{1}^{t_R}(p) =  \Pi^R_{Q_1}(p)-\Pi^R_{Q_4}(p)\, , \nonumber\\
&&M_{1}^{t}(p)     = M_{Q_1}(p) - M_{Q_4}(p) \, .
\label{selff}
\eea
Notice that we have canonically normalized the kinetic term of the top  in the limit in which the top decouples from the strong sector.
As in the case of the gauge correlators, 
$\Pi^{L,R}_{Q_{4,1}}$ and $M_{Q_{4,1}}$ can be written in a large-$N$ expansion as a sum over infinite resonances.
We have
\begin{equation}
\Pi_{Q_{4}}^L(p)=  \sum_n\frac{{|F^L_{Q^{(n)}_4}}|^2}{p^2+m^2_{Q^{(n)}_4}}\, ,\qquad
\Pi_{Q_1}^L(p)=
 \sum_n\frac{{|F^L_{Q^{(n)}_1}}|^2}{p^2+m^2_{Q^{(n)}_1}} \, ,
\end{equation}
and similarly for $\Pi_{Q_{4,1}}^R$ with the replacement $L\rightarrow R$,
 while
\begin{equation}
M_{Q_4}(p)=  \sum_n\frac{F_{ Q^{(n)}_4}^L  F^{R\, *}_{Q^{(n)}_4}  m_{Q^{(n)}_4}}{p^2+m^2_{Q^{(n)}_4}}\, ,\qquad
M_{Q_1}(p)= \sum_n\frac{F_{Q^{(n)}_1}^LF^{R\, *}_{Q^{(n)}_1} m_{Q^{(n)}_1}}{p^2+m^2_{Q^{(n)}_1}}\, .
\label{corrfermion}
\end{equation}
We denote with $Q^{(n)}_4$ and $Q^{(n)}_1$ the (color-triplet) vector-like  fermonic resonances with the SO(4) quantum numbers of  $\bf 4$ and $\bf 1$ respectively. The dimensionful parameters  
$F^{L,R}_{Q^{(n)}_i}$ are the equivalent of the decay constants of spin-1 resonances  but here for  fermions: they correspond to mixing mass-terms  between the top quark and the fermionic resonances. 
For convenience we define the mixing parameter 
\be
F^{L,R}_{Q^{(n)}_i}/m_{Q^{(n)}_i}\equiv \epsilon\, .
\label{mix}
 \ee

As with the gauge contribution, we want to calculate the correlators of \eq{selff}  considering only
 the minimal number of resonances necessary for the convergence of the Higgs potential \eq{potF}.
To obtain a finite result we need  that  at large momentum the form factors $\Pi_1^{t_{L,R}}$ fall off at least as $\Pi_1^{t_{L,R}}\rightarrow 1/p^6$.
This is equivalent to imposing three pairs of Weinberg sum-rules, $\lim_{p\rightarrow\infty}p^n\,\Pi_1^{t_{L,R}}(p)= 0$  $(n=0,2,4)$ that can be fulfilled  
with at least three vector-like resonances.
Our main interest, however, is in the calculation of the Higgs mass which, as we will now show,
 requires  less resonances to be finite.

Let us calculate the Higgs mass  in the approximation of small Higgs vacuum expectation value
$\langle s_h\rangle\ll 1$. 
 The potential \eq{potF} can be expanded as
\begin{equation}
V(h)= \alpha\ s^2_h-\beta\ s^2_hc^2_h+\cdots\, ,
\label{potential}
\end{equation} 
where $\alpha=O(\epsilon^2)$  and $\beta=O(\epsilon^4)$.
Notice that at least  two terms must be included in the expansion in order to  have realistic EWSB.
Indeed, for $\alpha< \beta$ and  $\beta\geqslant 0$ we have that the electroweak symmetry is broken:
\begin{equation}
 \langle s_h\rangle\equiv\frac{v}{f}\equiv\sqrt{\xi}\simeq\sqrt{\frac{\beta-\alpha}{2\beta}}\, ,
\label{vev}
\end{equation}
and the Higgs mass is given by
\begin{equation}
m_h^2\simeq \frac{8\beta}{f^2}\,  \langle s^2_hc^2_h\rangle\, ,
\label{mh2}
\end{equation}
with
\begin{equation}
\beta=N_c\int \frac{d^4p}{(2\pi)^4}\left[\frac{|M^t_1|^2}{p^2 \Pi_0^{t_L} \tilde\Pi_0^{t_R}}
          +\left(\frac{\Pi_{1}^{t_L}}{2\Pi_0^{t_L}}\right)^2+\left(\frac{\Pi_{1}^{t_R}}{\tilde\Pi_0^{t_R}}\right)^2\right]\, ,
\label{beta0}
\end{equation}
where $\tilde\Pi_0^{t_R}\equiv\Pi_0^{t_R}+\Pi_1^{t_R}$. 
If instead of 
an  expansion in  $\langle s_h^2\rangle$, we had performed an  expansion in $\epsilon^2$,
we would still have obtained Eqs.~(\ref{vev})-(\ref{beta0}), but with $\Pi_0^{t_{L}}=\tilde \Pi_0^{t_{R}}\simeq1$.
Let us from now on work in this limit, $\epsilon^2\ll 1$, 
 corresponding to small mixing between the top and the resonances;
we will see later that this is a good approximation for most of our calculations. 
From \eq{beta0} we can easily derive 
a lower-bound on the Higgs mass as a function of the lightest resonance mass.
This is based on the fact that  the three terms  in \eq{beta0} are positive, meaning that we can bound the Higgs mass using only the first one:
\begin{equation}
m_h^2\geq \frac{2 N_c}{\pi^2}\frac{m^2_t}{f^2}\int^\infty_0 \!\!\! dp\; p\, \left|\frac{M^t_1(p)}{M^t_1(0)}\right|^2\, ,
\label{cutoff}
\end{equation}
where we have used  the fact that the physical top mass is given by
\be
m_t= \frac{|M^t_1(0)|}{\sqrt{2\Pi_0^{t_L}(0) \tilde\Pi_0^{t_R}(0)}} \langle s_hc_h\rangle\, .
\label{mt}
\ee
The  convergence  of \eq{cutoff} requires the Weinberg sum-rule
$\lim_{p\rightarrow \infty}M^t_1(p)=0$.
This can be achieved with  just one resonance,  
\be
\left|\frac{M^t_1(p)}{M^t_1(0)}\right|=\frac{m_Q^2}{p^2+m_Q^2}\, ,
\ee
where $Q$ represents here  the lightest  resonance, that can either  be  a $\bf 4$ or a $\bf 1$
of SO(4), since
this procedure  does not depend on its quantum numbers.
We  then  have 
\begin{equation}
m_h^2\geq \frac{N_c}{\pi^2}\frac{m^2_t}{f^2}m_Q^2\, ,
\label{lowerbound}
\end{equation}
that provides an upper bound for the resonance mass:
\be
m_Q\lesssim 700\ \text{GeV}\left(\frac{m_h}{125\ \text{GeV}}\right)\left(\frac{160\ \text{GeV}}{m_t}\right)\left(\frac{f}{500\ \text{GeV}}\right)\, .
\label{upb}
\ee

\begin{figure}[t] 
\centering
   \includegraphics[width=0.5\textwidth]{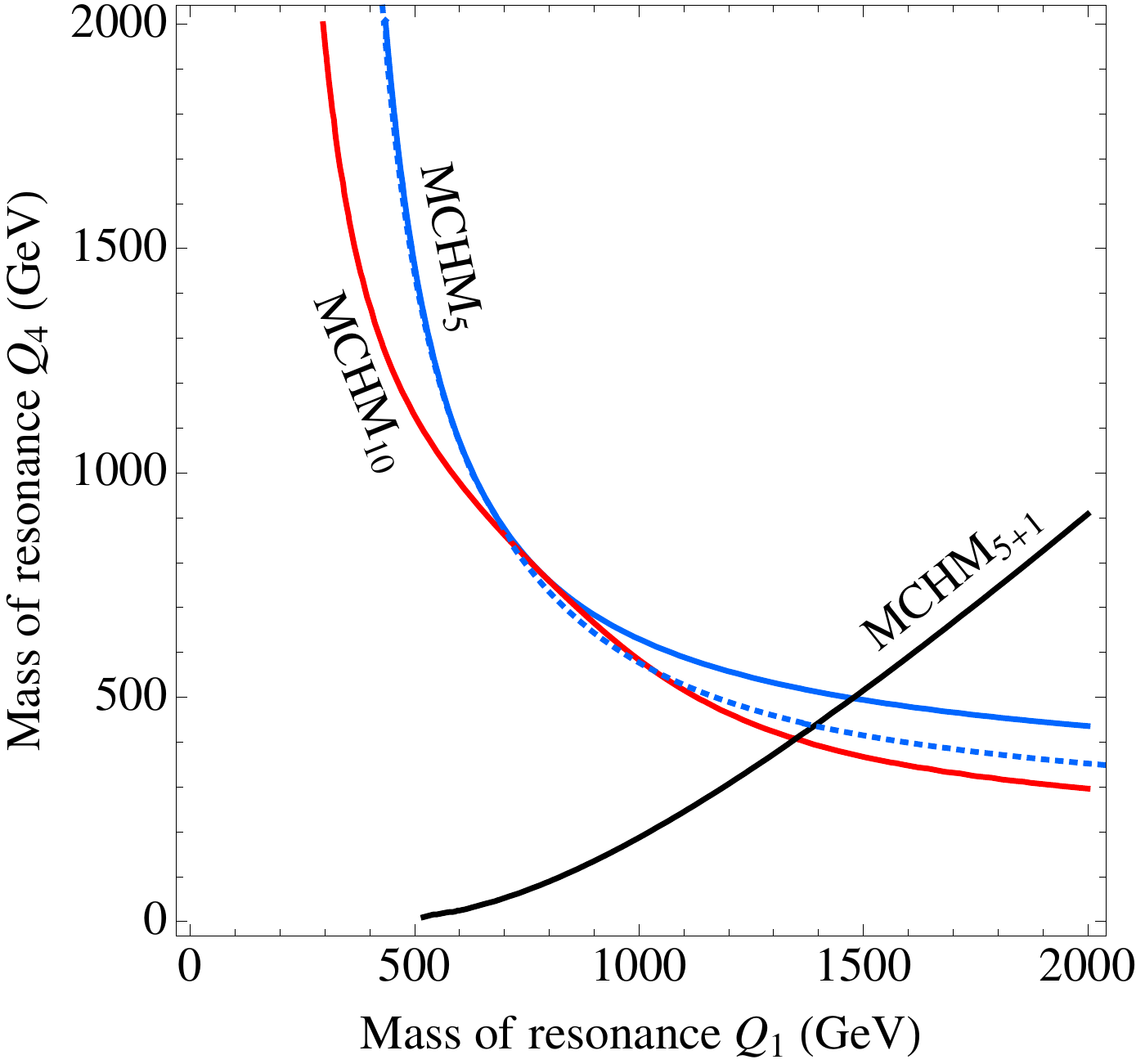}
 \caption{\footnotesize \emph{Masses of the two lightest fermion resonances  for  $m_h=125\GeV$ (taking $\xi=0.2$ and $m_t=160\GeV$ (the running top mass at  $\sim$ TeV)).  In blue we plot the  $MCHM_5$  result; the solid line corresponds to \eq{mh22}  calculated in the approximation $\epsilon^2\ll 1$,
while the dashed line is the exact result (always $\Delta F^2=0$).
In solid red we plot the   result for the $MCHM_{10}$ ($\epsilon^2\ll 1$ and $\Delta F^2=0$)  
with   ${Q_1}\rightarrow  {Q_6}$.
The black solid line is for ${\bf r_L=5}$ and ${\bf r_R=1}$ (denoted MCHM$_{5+1}$), fixing for illustration $F_{Q_1}^L=\sqrt{2}\tilde{F}_{Q_1}^R$.}}
 \label{figureTwoResonances}
\end{figure}

To obtain a convergent result for the Higgs mass from the full top-quark contribution of \eq{beta0},
we must  impose the  two pairs of Weinberg sum-rules,  $\lim_{p\rightarrow\infty}p^n\Pi_1^{t_{L,R}}(p)= 0$ $(n=0,2)$,
that require  at least two resonances, $Q^{(1)}_1\equiv Q_1$ and $Q^{(4)}_1\equiv Q_4$.
We obtain
\bea
\Pi_1^{t_{L,R}}&=&|F^{L,R}_{Q_4}|^2\frac{(m^2_{Q_4}-m^2_{Q_1})}{(p^2+m_{Q_4}^2)(p^2+m_{Q_1}^2)}\ ,\nonumber\\
M^t_1(p)&=&|F^L_{Q_4} F^{R\, *}_{Q_4}|\frac{ m_{Q_4} m_{Q_1}(m_{Q_4}-m_{Q_1} e^{i\theta})}{(p^2+m_{Q_4}^2)(p^2+m_{Q_1}^2)}\left(1+
\frac{p^2}{m_{Q_4}m_{Q_1}}\frac{m_{Q_1}-m_{Q_4} e^{i\theta}}{ m_{Q_4}-m_{Q_1} e^{i\theta}}\right)\, ,
\label{correl}
\eea
where we have defined $F^L_{Q_4}F^{R\, *}_{Q_4}=e^{i\theta}|F^L_{Q_4}F^{R\, *}_{Q_4}|$ and set by  a field redefinition $F^L_{Q_1}F^R_{Q_1}$ to be real. 
\eq{correl} together with \eq{mt} gives~\footnote{A similar expression has also been obtained in the context of deconstructed MCHM \cite{Panico:2011pw}.}
\begin{equation}
m_h^2\simeq \frac{N_c}{\pi^2}\left[\frac{m^2_t}{f^2}\frac{m_{Q_4}^2 m^2_{Q_1}}{m_{Q_1}^2-m^2_{Q_4}} \log\left(\frac{m^2_{Q_1}}{m^2_{Q_4}}\right)+\frac{(\Delta F^2)^2}{4 f^2}\langle s^2_h c^2_h\rangle\left(\frac{1}{2}\, \frac{m_{Q_4}^2+ m^2_{Q_1}}{m_{Q_1}^2-m^2_{Q_4}} \log\left(\frac{m^2_{Q_1}}{m^2_{Q_4}}\right)-1\right)
\right]\, ,
\label{mh22}
\end{equation}
where $\Delta F^2=|F^{L}_{Q_4}|^2-2|F^{R}_{Q_4}|^2$.
 It is easy to see that the second term in \eq{mh22} is always
positive and that  the first term minimizes  for   $m_{Q_4}\rightarrow m_{Q_1}$ where the Higgs mass
saturates the lower-bound \eq{lowerbound}.  
It is also important to notice that, considering only the top contributions  to the  Higgs potential, 
one obtains that   $\alpha$ in \eq{potential}  is proportional to $\Delta F^2$, 
meaning that  the condition  $\alpha<\beta$ requires  small   values for $\Delta F^2$.   
 In this limit, the Higgs mass comes entirely   from the first term of \eq{mh22}. 
  In Figure~\ref{figureTwoResonances}   we show the value of the two lightest resonance masses
for a Higgs mass $m_h=125$ GeV.     
One can see that  there is always a light state with a mass roughly  between 500 GeV and 700 GeV.
Now, since light resonances imply  large values of $\epsilon$ (see \eq{mix}), one could worry about the validity of our approximation $\epsilon^2\ll 1$.     In Figure~\ref{figureTwoResonances}  we show with a dashed blue line
  the result obtained 
  without taking the small $\epsilon^2$ limit.    As can be appreciated, 
    the differences are small  and the approximation $\epsilon^2\ll 1$ always gives a more conservative upper-bound 
on the resonance masses. 
We note however that  in the exact result
   the masses of the lightest fermionic resonances  differ from
 $m_{Q_{1,4}}$  due to  the sizable mixings  with $t_{L,R}$.
Therefore not only a light Higgs implies light fermionic resonances, but also
a sizable degree of compositeness of the top.

A very similar model to the MCHM$_5$ is the MCHM$_{10}$ \cite{Contino:2006qr}, 
in  which the left-handed and right-handed top quarks are   embedded     into spurions 
in  the $\bf 10$ representation of SO(5).
In this model Eqs.~(\ref{potential})-(\ref{mh2}) also hold, and $\beta$  is given by~\footnote{As in  Ref.~\cite{Contino:2006qr},  we are  not considering
invariants formed  by contracting the spurions with the Levi-Civita tensor (see Appendix B). 
These  invariants can be eliminated  by imposing  extra parities in the strong sector, along the lines
of the models in Ref.~\cite{Mrazek:2011iu}.}
\begin{equation}
\beta= N_c\int \frac{d^4p}{(2\pi)^4}\left[\frac{|M^t_1|^2}{8p^2 \Pi_0^{t_L} \Pi_0^{t_R}}
          +\left(\frac{3\Pi_{1}^{t_L}}{4\Pi_0^{t_L}}\right)^2+\left(\frac{\Pi_{1}^{t_R}}{4\Pi_0^{t_R}}\right)^2\right]\, ,
\label{beta1}
\end{equation}
where now $m_t= {\langle s_hc_h\rangle |M^t_1(0)|}/{\sqrt{16 \Pi_0^{t_L}(0) \Pi_0^{t_R}(0)}}$
and  the  correlators  are
  the same as \eq{correl} but with the replacement $Q_1\rightarrow Q_6$, since  $\bf 10=4\oplus 6$ under SO(4).
For the Higgs mass   we obtain \eq{mh22}      with  the replacement
$(\Delta F^2)^2\rightarrow (\Delta F^2)^2+|F^{L}_{Q_4}F^{R}_{Q_4}|^2$ where
now $\Delta F^2=(3|F^{L}_{Q_4}|^2- |F^{R}_{Q_4}|^2)/2$.
In this model as well, $\langle s_h\rangle\ll 1$  requires $\Delta F^2$ small, 
implying   that 
  the Higgs mass   in the  MCHM$_{10}$ is always  larger than that in
the MCHM$_5$. This is shown in Figure~\ref{figureTwoResonances}.

We then conclude that a light Higgs  requires a fermionic resonance mass  smaller  than the vector one $m_\rho$ that, due to  constraints from electroweak precision tests  \cite{Giudice:2007fh},
must  lie above  $2-3$ TeV~\footnote{At the one-loop level the fermionic resonances can contribute to electroweak observables.
The precise magnitude of these contributions depend on the details of the composite models and will not be discussed here.}. 
A possible natural explanation for this  mass splitting  in the resonance spectrum
can be found in  holographic models  \cite{Contino:2006qr}.
In these models one finds that 
 in the limit in which   $t_L$ (or $t_R$)  
has large mixings   with the composite sector, $\epsilon\sim1$, 
    extra  massive resonances, called custodians, become light and            complete   with  $t_L$
(or $t_R$)   a full  SO(5) multiplet.
This phenomenon can be understood in the following way.
The mixing of the  top with  the strong-sector resonances depends on
the dimension 
of the  chiral  fermionic operator to which the top couples to  (see \eq{mixingsms} in Appendix~A).
This mixing  becomes more sizable  as we decrease the dimension  of this operator;
when this dimension approaches the lower- bound  3/2, the fermionic  operator becomes a  free field  that  decouples from
the strong sector and then it mass tends to zero.  
In the MCHM$_5$   for a  $t_L$ with sizable mixings $\epsilon$,  the light resonances are
those states that complete with $t_L$ a  $\bf 5$ of SO(5) \cite{Contino:2006qr}: 
colored states with quantum numbers 
 $\bf 1_{2/3}$ and  $\bf 2_{7/6}$ under the SM SU(2)$_L\times$U(1)$_Y$.
Similarly, for  a $t_R$ with large mixing $\epsilon$,  these states  correspond to 
two weak-doublets,  $\bf 2_{1/6}$ and $\bf 2_{7/6}$,  forming together a    $\bf 4$-plet of SO(4).
   This connection between  large mixings and light resonances 
has  also been  studied in simple  models of partly composite top \cite{Pomarol:2008bh}
or, more recently, via deconstructed versions of the MCHM \cite{Panico:2011pw,DeCurtis:2011yx}.

\subsection{Extension to other  representations}

It is important  to analyze whether the  fact that a  light Higgs implies fermionic resonances below the TeV
 goes beyond the minimal realizations MCHM$_{5,10}$.  
 In this section we will consider the SO(5)/SO(4) model with 
 the left-handed and right-handed top quarks
embedded respectively  in generic  representations  $\bf r_L$ and $\bf r_R$ of SO(5).
The generalization of \eq{efflag}  
 can be written in this case as
\bea
{\cal L}_{\rm eff}&=&
   \bar b_L \pslash 
   \left(
 \sum_{i}  c_h^{i}\, \Pi^{b_L}_{i}(p)
 \right)
 b_L+
  \bar t_L \pslash
  \left( 
  \sum_{i}  c_h^{i}\, \Pi^{t_L}_{i}(p)
\right)
t_L+
  \bar t_R \pslash 
  \left(
  \sum_{i}  {c_h^{i}}\, \Pi^{t_R}_{i}(p)
  \right)
  t_R\nonumber\\
& +&
 \big( s_h^{1+2m} c_h^{n}\ \bar t_L  M^t_{1}(p) t_R
+h.c.\big)\, ,
 \label{efflag2}
\eea
where    $i,m,n$ are positive integers.
Some examples   are given in Appendix B.
In \eq{efflag2} we have included a contribution from $b_L$ that can also  be  present 
for some embeddings.
In order to guarantee the absence of flavor-violation, we have assumed that
 only one operator is responsible for the fermion masses \cite{Mrazek:2011iu}.
In this section we will only consider models with $m=0$ which, as we will show in the next section, can be preferred by the experimental data if  a light    Higgs   with SM-like couplings were confirmed.   

One of the  most important  differences between  the MCHM$_5$ and these generic models
is that the Higgs mass squared  can arise  at $O(\epsilon^2)$  ({\it i.e.}, $m_h^2\sim \int d^4p\,\Pi_{i>0}$), 
instead of $O(\epsilon^4)$~\footnote{See Ref.~\cite{Mrazek:2011iu}
 where models with this property  were first proposed for the  SO(6)/SO(4) coset.}.
This can happen when at least two terms with different powers of $c_h$ 
appear in the  effective kinetic terms of  $t_{L,R},b_L$  (first line of \eq{efflag2}),
  generating  then a potential $V(h)=\alpha c^i_h+\beta c^j_h+\cdots$  $(i\not=j)$
with both $\alpha$ and $\beta$ of $O(\epsilon^2)$. 
The interesting feature of these scenarios  is that a nontrivial minimum 
with $\langle s_h\rangle\ll 1$ can be accommodated more easily
than in the MCHM$_5$ where \eq{vev} requires $\alpha<\beta$  and  $O(\epsilon^2)$-terms 
 to be smaller than $O(\epsilon^4)$-terms.

Although this type of models can be thought to be more natural than  the MCHM$_5$,
they generically predict a heavier   Higgs  and are therefore disfavored by
the present data that hints towards a light Higgs.    
To quantify this, let us consider  the case   $\bf r_L=14$ and $\bf r_R=1$.
  The Higgs potential has the same form as \eq{potential}  where now
\begin{equation}
\beta\simeq 2 N_c\int \frac{d^4p}{(2\pi)^4} \frac{\Pi_{4}^{t_L}}{\Pi_0^{t_L}}\, ,
\end{equation}
with  $\Pi_4^{t_L}=5\Pi^L_{Q_1}/4-2\Pi^L_{Q_4}+3\Pi^L_{Q_9}/4$ (recall that under SO(4), $\bf 14=9\oplus4\oplus1$ 
and there are therefore  three types of resonances $Q^{(n)}_{1,4,9}$ entering in $\Pi^{t_L}_4$; see Appendix B).
To obtain a finite $\beta$  we need to consider  at least   3 resonances,
 $Q^{(1)}_1\equiv Q_1$, $Q^{(1)}_4\equiv Q_4$ and
$Q^{(1)}_9\equiv Q_9$ that, after imposing the Weinberg sum-rules, gives (for  $m^2_{Q_4}\not= m^2_{Q_9}$)
\be
\Pi_{4}^{t_L}=\frac{5}{4}|F^L_{Q_1}|^2\frac{ (m^2_{Q_4}-m^2_{Q_1})(m^2_{Q_9}-m^2_{Q_1}) }{(p^2+m^2_{Q_1})(p^2+m^2_{Q_4})(p^2+m^2_{Q_9})}\, ,
\label{cor14}
\ee
and  a Higgs mass
\be
m^2_h\simeq  \frac{N_c}{\pi^2}\frac{5 |F^L_{Q_1}|^2}{4 f^2}
\left[m^2_{Q_1}\log\left(\frac{m^2_{Q_1}}{m^2_{Q_9}}\right)+
\frac{m_{Q_4}^2 (m^2_{Q_9}-m^2_{Q_1})}{m_{Q_9}^2-m^2_{Q_4}} \log\left(\frac{m^2_{Q_9}}{m^2_{Q_4}}\right)\right]\, .
 \ee
 The value of $F^L_{Q_1}$  is related to the top mass: 
 \be
m_t=\frac{|F^L_{Q_1} \tilde F^{R\, *}_{Q_1}|}{m_{Q_1}}\langle c_hs_h\rangle\,  ,
\label{mt2}
 \ee
where $\tilde F^{R}_{Q_1}$ is the mixing of $Q_1$ with  $t_R$.
Therefore  $F^L_{Q_1}$ has a lower bound for a given $\tilde F^{R}_{Q_1}$ and the smallest value of  $m_h$ is achieved for the 
 largest value of  $\tilde F^R_{Q_1}$, or equivalently,  when $t_R$  is fully composite.
 In this case it is simpler to  start  from the beginning assuming that $t_R$ appears as a 
massless composite state from the strong sector.  The calculation for the Higgs mass is then quite easy. The form factor $\Pi^{t_L}_4$ has to have a massless pole, corresponding to $t_R$, with a residue given by $m_t^2/\langle s_h^2c_h^2\rangle$. The Weinberg sum-rules require in this case only two massive resonances, for instance $Q_9$ and $Q_4$,  leading to 
\be
\Pi_4^{t_L}=\frac{m^2_t}{\langle s^2_h c^2_h\rangle}\frac{m_{Q_9}^2m^2_{Q_4}} {p^2(p^2+m^2_{Q_9})(p^2+m^2_{Q_4})}\, ,
\ee
and then a Higgs mass
\begin{equation}
m_h^2\simeq \frac{N_c}{\pi^2}\left[\frac{m^2_t}{f^2}\frac{m_{Q_4}^2 m^2_{Q_9}}{m_{Q_4}^2-m^2_{Q_9}} \log\left(\frac{m^2_{Q_4}}{m^2_{Q_9}}\right)
\right]\, ,
\label{mh2222}
\end{equation}
which is the same expression as for the MCHM$_5$, \eq{mh22},  with $\Delta F^2=0$ and $Q_9$ playing the role of $Q_1$. Thus, a Higgs mass of order $125$ GeV, implies again light resonances. Nevertheless, 
in this case, 
we cannot rely on arguments  based on 
holographic models~\cite{Contino:2006qr,Pomarol:2008bh} to naturally expect  light resonances.
As discussed in the previous section,
only in the limit of sizable mixing between the top and the strong sector,  holographic models  predict light resonances.
 In  models with $\bf r_R=1$ there are not custodians  associated to $t_R$.
Furthermore,  when  $t_R$ is fully composite,  
 the  $t_L$ 
must have a small mixing to the strong sector to predict the right top mass  and therefore its 
  custodians are not expected to be light.
We could alternatively  reduce the Higgs mass to $\sim 125$ GeV   by tuning, either by demanding a larger value of $f$, or 
 by requiring   cancellations in  $\Pi_{4}^{t_L}$.  This latter possibility can be realized, for instance, 
in the presence of a certain degree of 
degeneracy between the resonances of different SO(4)-multiplets (for example in the limit $m _{Q_9}\rightarrow m_{Q_1}$
in \eq{cor14}).

To further explore  the relation between the Higgs mass and the resonance mass
in  generic SO(5)/SO(4) models,
we  would like to derive  here a generalization of the   Higgs mass lower-bound \eq{lowerbound}
 under certain reasonable assumptions.
To do this we  use the fact that 
in any   composite Higgs model,  the Higgs mass   must  
receive at least the   model-independent contribution 
arising  from the top-mass form-factor $M^t_1(p) $ that  is always sizable since
  $M^t_1(0)\propto m_t$. 
From the last term in the lagrangian \eq{efflag2}, this  contributes to the Higgs potential
\be
\Delta V_t(h)=-\beta_{t}\ s^2_hc^{2n}_h\, ,
\label{DeltaVt}
\ee
where
\be
\beta_{t}=\frac{N_c}{8 \pi^2}\frac{m_t^2}{s^2_h c_h^{2n}} m_Q^2\ , \ \ \ \ \ \ 
m^2_Q\equiv 2\int_0^\infty p\,dp \left|\frac{M^t_1(p)}{M_1^t(0)}\right|^2\, .
\label{boundbeta}
\ee
Realistic EWSB requires $v/f\ll 1$, a limit that can be  achieved in two different ways.
The first possibility is to  take a  large value of $n$. Indeed, from the minimization
of $\Delta V_{t}(h)$, we get
$\langle s^2_h\rangle=1/(1+n)$ that tends to zero for $n\rightarrow\infty$.
In this minimum, we find
\begin{equation}\label{mhnh}
m_h^2\simeq  \frac{N_c}{2 \pi^2}\frac{m_t^2}{v^2} m_Q^2\, ,
\end{equation}
valid for any value  of $n$.  \eq{mhnh}
corresponds to a  Higgs mass  a factor
 $\sqrt{2}f/v$ larger than  \eq{lowerbound},
 implying  then resonance masses   even lighter than in \eq{upb}.
 
The other possibility to have $v/f\ll 1$ is by adjusting 
parameters in the potential, as it was the case in all models discussed so far.
For example, consider adding to $\Delta V_t(h)$
the term $\Delta V'=\alpha  c^{m'}_h$.
A small $v/f$ can now be obtained  if   $\alpha\simeq -2 \beta_t/ m'$,
giving in this  limit~\footnote{
The other interesting possibility is to   add
 $\Delta V'=\alpha  s_h^2c^{ m'}_h$ that, instead of  predicting  \eq{mhaa}, gives 
$m_h^2\simeq \frac{4\beta_t v^2}{f^4}(2n-m')$,
leading to a Higgs mass even larger than the lower-bound \eq{lowerbound2}.}
\be
m_h^2=\frac{4\beta_t\, v^2}{f^4}\left[1-\frac{m'}{2}+2n  +\left(1-\frac{m'}{2}+n(1+m'-3n)\right)\frac{v^2}{f^2}+\cdots \right]\, . 
\label{mhaa}
\ee
From \eq{mhaa} we can  write  a  lower bound  valid for any value of $n$ and $m'$ such that  $m'\not=4n+2$:   
\begin{equation}
m_h^2\gtrsim \frac{N_c}{4 \pi^2}\frac{m^2_t}{f^2}m_Q^2\, ,
\label{lowerbound2}
\end{equation}
that is a factor $4$ smaller than \eq{lowerbound} and, for a given Higgs mass, allows resonance masses $m_Q$ a factor $2$ larger.
For the particular case $m'=4n+2$, on the other hand,
 the leading  term in   \eq{mhaa} 
 vanishes and the Higgs mass squared is then proportional to $v^4/f^4$. We then have
 \begin{equation}
 \label{MH2}
m_h^2\simeq n(1+n) \frac{N_c}{2 \pi^2}\frac{m_t^2}{f^2}\frac{v^2}{f^2} \ m_Q^2\, ,
\end{equation}
allowing for even   larger resonance masses for $n$ small. 

We finish this analysis by pointing  out a possible caveat in  the 
argument that has led to 
the Higgs mass lower-bounds discussed above.
It could well be  that  the contribution to the Higgs potential coming from $\beta_t$ is
  cancelled  by other terms in the potential in order to obtain  realistic EWSB,
 not giving then  any contribution to the Higgs mass.
This occurs, for example,
in models with 
the embeddings $\bf r_L= 10\, ,\ r_R= 5$ or viceversa~\footnote{This type of models were first  considered in Ref.~\cite{Frigerio:2012uc} for the SO(6)/SO(5) coset.},  
and $\bf r_L=\bf 5\, ,r_R= 1$.  
Let us  consider  this latter case.
The potential, in the $\epsilon^2$-expansion, is given by  $V(h)= \alpha s^2_h +\beta s^4_h+\cdots$ where
\bea
\alpha&=&-\beta_t|_{n=0}-2 N_c\int \frac{d^4p}{(2\pi)^4}\left(\frac{\Pi_{1}^{t_L}}{2\Pi_0^{t_L}}\right)\, ,
\label{alpha2}
\\
\beta&=& 
N_c\int \frac{d^4p}{(2\pi)^4}\left(\frac{\Pi_{1}^{t_L}}{2\Pi_0^{t_L}}\right)^2\, .
\label{alpha51}
\eea
Notice that
the  contribution from the top-mass form-factor, $\beta_t$, enters now in $\alpha$ and not in $\beta$.
The minimization  of the potential gives $\langle s_h\rangle=\sqrt{-\alpha/(2\beta)}$ and 
the Higgs mass is given by  $m_h^2\simeq {8 \beta}\langle s^2_hc^2_h\rangle/{f^2}$,
that    does not receive any contribution from $\beta_t$.
The Higgs mass can then be  much smaller than in the models discussed above.  
For two resonances, we have
\be
m_h^2\simeq \frac{N_c }{4 \pi^2}\frac{|F^L_{Q_1}|^4\langle s_h^2c_h^2\rangle}{f^2}\left(\frac{1}{2}\frac{m_{Q_4}^2+m_{Q_1}^2}{m_{Q_4}^2-m_{Q_1}^2}\log \left(\frac{m_{Q_4}^2}{m_{Q_1}^2}\right)-1\right)\, ,
\ee
that is of $O(\epsilon^4)$ and can thus be quite small
if  $F^L_{Q_1}$ is small.  As in the 
$\bf r_L=14$, $\bf r_R=1$ case,
the smallest value of $m_h$ is found  
for maximal mixing $\tilde{F}^R_{Q_1}$ of $t_R$ with the composite resonances, i.e. 
a $t_R$  fully composite~\footnote{The limit of a fully composite $t_R$ is also necessary if $\alpha$ of \eq{alpha2} has to be   tuned
with the gauge contribution \eq{gaugecon1} to guarantee $\langle s_h\rangle\ll 1$.}.
Again, it is simpler to start with  $t_R$ as a massless resonance arising from  the strong sector.
We then find that 
 one extra massive resonance is enough   to satisfy the convergence of the integral in \eq{alpha51}.
We have  
$\Pi_1^{t_L}(p)=(m^2_t/\langle s_h^2\rangle)m^2_{Q_4}/(p^2 (p^2+m^2_{Q_4}))$ leading to a Higgs mass
\be
m^2_h \simeq \frac{N_c}{8\pi^2}\frac{m^4_t \langle c^2_h\rangle}{v^2}\left[\log \left(\frac{m_{Q_4}^2+m^2_t}{m^2_t}\right)-\frac{m_{Q_4}^2}{m^2_{Q_4}+m^2_t}\right]\, .
\label{mh2f}
\ee
For $m_{Q_4}\simeq 3$ TeV, the Higgs mass \eq{mh2f}  can be as small as $40$ GeV.
Larger values of $m_h$ imply larger values of $F_{Q_1}^L$, meaning that
 $m_h\sim $125 GeV can be obtained without light fermionic resonances as we show in Figure~\ref{figureTwoResonances}.
In this case, however, it is important to notice that extra contributions are needed to reduce $\alpha$  in order to have $\langle s_h\rangle\ll 1$.

\section{Higgs couplings to SM fermions}
\label{sec:HiggsCouplings}

\begin{figure}[t] 
\centering
   \includegraphics[width=0.5\textwidth]{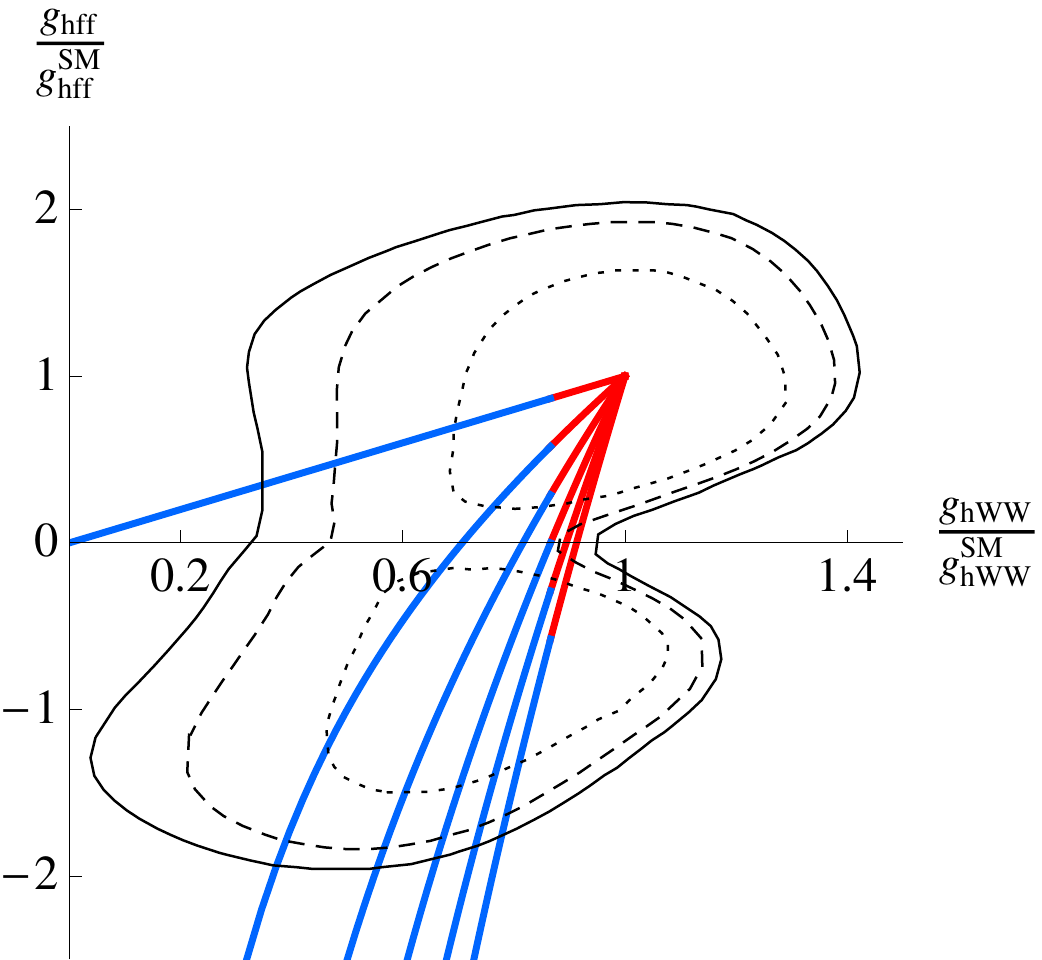}
 \caption{\footnotesize \emph{Predictions  of a generic MCHM in the 
 ($g_{hff}/g^{\rm SM}_{hff}$, $g_{hWW}/g^{\rm SM}_{hWW}$)-plane.  
 The different curves  corresponds to different values of  $n$,
  going downwards from n=0 to $n=5$. The red part of the curves is for $0<\xi<0.25$ and  the blue one for $0.25<\xi<1$. 
    The contours are the $68\%$, $95\%$ and $99\%$ CL
 for a 125 GeV Higgs     as  obtained in Ref.~\cite{Azatov:2012bz} from the CMS data.}}
 \label{figureghff}
\end{figure}

In  composite Higgs  models  
 the Higgs couplings  to fermions  generically deviate from their SM values \cite{Giudice:2007fh}.
For the  SO(5)/SO(4) model, the Higgs couplings to  the SM  fermions
can be parametrized by  \eq{efflag2}. 
At low-energies  $p\ll m_{Q_i}$ and in the limit   $\epsilon\ll 1$, 
the Higgs couplings  reduce,  for the case of a generic SM fermion $f_{L,R}$, to
\be
 {\cal L}_{\rm eff}\simeq \bar f_L  M_1^f(0) f_R s_h^{1+2m}c_h^{n}+h.c. \equiv
\bar f_L f_R m_f(h)+h.c\, .
\ee
From this  we can obtain the $hff$ coupling
\cite{Giudice:2007fh}:
\be
\frac{g_{hff}}{g_{hff}^{\rm SM}}=\frac{2 m_W(h)}{g m_f(h)}\frac{\partial  m_f(h)}{\partial h}=\frac{1+2m-(1+2m+n)\xi}{\sqrt{1-\xi}}\, ,
\label{deformation}
\ee
where   we have used that $m_W(h)=gs_h/2$  \cite{Agashe:2004rs}
and written  the SM $hff$ coupling as a function of the physical $W$ and fermion mass, $g_{hff}^{\rm SM}=g m_f/(2 m_W)$.
 For $m\not=0$, \eq{deformation} gives deviations  of order one from the SM expectations, even in the
limit $\xi\rightarrow 1$. 
For this reason, we will concentrate on the $m=0$ case.
In Figure~\ref{figureghff}  we show, for  $m_h\simeq 125$ GeV and assuming that all fermions couple in the same way, 
the  $68\%$, $95\%$ and $99\%$ CL contours for $g_{hff}$ and $g_{hWW}$  extracted from the most recent CMS data
\cite{Azatov:2012bz}  (see also  \cite{Espinosa:2012ir} for similar analyses).
We have used that the $hWW$ coupling  in the SO(5)/SO(4) model is  given by $g_{hWW}=\sqrt{1-\xi}\, g^{\rm SM}_{hWW}$ \cite{Giudice:2007fh}.
Notice that models with large $n$  predict  negative values of $g_{hff}$, 
and then lie
in the lower half-plane of Figure~\ref{figureghff}, which
is also experimentally favored. 

An interesting prediction for   models where the Higgs potential is dominated by  the top-mass term $\Delta V_t$, \eq{DeltaVt}, is that the Higgs is necessarily  fermiophobic, corresponding to the horizontal line $g_{hff}=0$ in 
Figure~\ref{figureghff}.  This is due to the fact that the minimization condition of the potential implies  
 $\partial  m_t(h)/\partial h=0$ and then, from \eq{deformation}, $g_{hff}=0$.

\section{Discussion and Conclusions}
\label{sec:Conclusions}

Using the  Weinberg sum-rules in the large-$N$ limit, that require the proper convergence of  correlators at large momentum,
and using the assumption that these correlators are dominated by the lowest-mass resonances, 
we have computed  in the MCHM the mass of the Higgs 
as a function  of the fermionic resonance masses. 
This has allowed us to show that  a light Higgs implies generically light fermionic resonances.

\begin{figure}[t] 
\centering
   \includegraphics[width=0.6\textwidth]{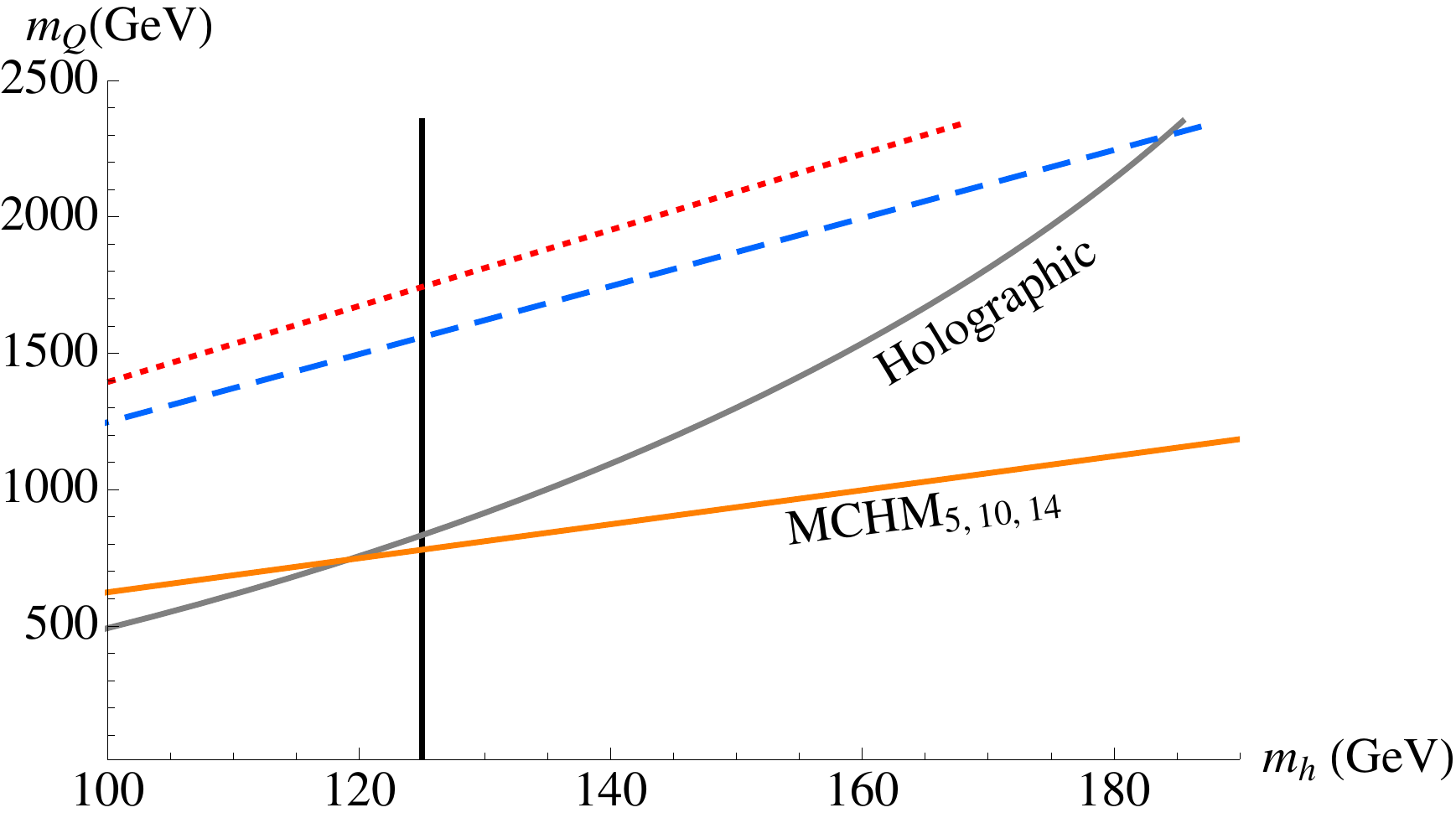}
 \caption{\footnotesize \emph{Upper bound for the mass of the lightest fermionic resonance in various composite Higgs models, for $\xi=0.2$ and $m_t=160\GeV$ (the running top mass at  $\sim$ TeV). The orange line corresponds to  \eq{lowerbound} that is the upper bound for the MCHM$_{5,10,14}$.
  The   bound  \eq{lowerbound2} is shown in dashed blue, 
while  \eq{MH2}  is shown in dotted red for $n=1$ and $m'=6$. A comparison with the Holographic MCHM$_{5}$ model of Ref. \cite{Contino:2006qr} is shown in grey.  The vertical line is for $m_h=125\GeV$.}}
 \label{figureHiggsMass}
\end{figure}

In Figure~\ref{figureHiggsMass} 
we give a brief summary  of the upper bounds  on $m_Q$, the lightest  fermionic resonance mass, 
obtained for different classes of MCHM.
With  an orange line  we show the upper bound  derived  for the 
 MCHM$_5$ \cite{Agashe:2004rs} and  MCHM$_{10}$ \cite{Contino:2006qr} where
 $m^2_h$  is  of $O(\epsilon^4)$.   We can see that  a light Higgs, $m_h\sim 125$ GeV, implies in this type of models resonances below the TeV. 
This upper bound on $m_Q$
also applies  for certain models where   $m^2_h=O(\epsilon^2)$, as  
 for example those with ${\bf r_L=14}$ and ${\bf r_R=1}$ (MCHM$_{14}$).
The smallest Higgs mass  
is obtained in this case  for  a fully composite $t_R$, giving   the same upper bound  on $m_Q$
as in the  MCHM$_5$. 

In  Figure~\ref{figureHiggsMass} we also
show 
in dashed blue  the  conservative upper-bound  \eq{lowerbound2},
derived under the assumption that the Higgs mass receives contributions from    the top form-factor \eq{DeltaVt}.
As we have shown,  however, this bound can be evaded in
very particular cases; for example,
  whenever 
the cancellation needed to reduce the quadratic term in the Higgs potential in order to achieve realistic EWSB, also implies, 
by accident,  a cancellation of the quartic. 
This can occur at the leading order in $v^2/f^2$, relaxing the bound on $m_Q$ to \eq{MH2} 
 (dotted red line in Figure~\ref{figureHiggsMass} for $n=1$), or  more drastically to all orders in $v^2/f^2$
 as in models with  ${\bf r_L=5}$ and ${\bf r_R=1}$.
In this latter case $m_h=125\GeV$  can be obtained even for resonance masses well above $1 \TeV$, as shown
 by the black line in Figure~\ref{figureTwoResonances}.
 
 We have studied models based on the minimal coset SO(5)/SO(4), but this method can be easily 
 extended to larger cosets \cite{Mrazek:2011iu,Gripaios:2009pe}.

 We can then conclude that,  with the exception  of some particular models, 
   a light Higgs $\sim125$ GeV
requires  generically the presence of  fermionic resonances below 
$m_\rho\sim 2-3$ TeV that should be possible to discover  at the LHC \cite{Contino:2008hi}.
This also requires that the top should have 
 a sizable degree of compositeness with important collider implications
\cite{Pomarol:2008bh,Lillie:2007hd}.

\vskip1cm
{\bf Note added:}
The 4th of July  2012,  LHC searches  \cite{cern} have confirmed  the presence of a Higgs-like state with a mass around 125 GeV.
An update of Fig.~\ref{figureghff} including the new experimental data can be found in Ref.~\cite{Montull:2012ik}.

\section*{Acknowledgments}
We  thank Roberto Contino  for providing  us  the fits of Figure~\ref{figureghff}.
The work of AP was partly supported by the projects FPA2011-25948, 2009SGR894 and ICREA Academia Program. 

\section*{Appendix A}

In this Appendix we derive   the relation between the top-quark form factors $\Pi^{t_{L,R}}_{0,1}$, $M^t_1$ and the
correlators  of the strong sector $\Pi^{L,R}_{Q_i}$,$M_{Q_i}$, following Refs.~\cite{{Agashe:2004rs},{Contino:2006qr}}. 
We do this for the case $\bf r_{L,R}=5$, and summarize the results for other interesting representations in Appendix B; generalization to other cases is straightforward.
 
Let us start defining the external sources $\Psi_{L,R}$ that couple to the operators ${\cal O}_{L,R}$ of the new strong sector:
\be
{\cal L}_{strong}+\bar \Psi_L {\cal O}_R+\bar \Psi_R {\cal O}_L\, .
\label{mixingsms}
\ee
The sources $\Psi_{L,R}$ are in complete SO(5) multiplets since the  lagrangian of the strong sector ${\cal L}_{strong}$, defined 
in the ultra-violet, is assumed to be SO(5) invariant. In our particular example $\Psi_{L,R}\in \bf 5$.
At low-energies, however, the SO(5) symmetry is  assumed to be spontaneously broken down to SO(4),
and therefore   the partition function  $\cal Z$
must be written as a functional 
 of the external sources  decomposed, now, into SO(4) multiplets $Q^{(r)}_{L,R}$,
 where $r$ labels the SO(4) representation.
In our case, since $\bf 5=4\oplus1$ under SO(4), we write
$\Psi_{L,R}=(Q_{L,R}^{(4)}, Q_{L,R}^{(1)})$.
Integrating over the strong sector, we find the partition function
    ${\cal Z}[Q^{(r)}_{L,R}]=e^{-\int {\cal L}_{\rm eff}}$ where, at the quadratic order,  we  have
\begin{equation}
{\cal L}_{\rm eff}=
  \bar Q_L^{(4)}  \pslash\,   \Pi^L_{Q_4}(p) Q_L^{(4)}+
  \bar Q_L^{(1)}  \pslash\,   \Pi^L_{Q_1}(p) Q_L^{(1)}+(L\rightarrow R)+
  \bar Q_L^{(4)} M_{Q_4}(p) Q_R^{(4)}+
  \bar Q_L^{(1)}  M_{Q_1}(p) Q_R^{(1)}+h.c.\, .
   \label{efflag3}
\end{equation}
The $\Pi_{Q_r}^{L}$ are related to the  two-point functions of   operators of the strong sector according to
\be
\pslash\,\Pi^L_{Q_r}(p)=\frac{\delta^2 {\cal L}_{\rm eff}}{\delta  \bar Q_L^{(r)}\delta Q_L^{(r)}}=\langle  {\cal O}_R^{(r)}(p)
\bar {\cal O}^{(r)}_R(-p)\rangle\, .
\ee
With the  use of the SO(5)/SO(4) Goldstones \cite{Agashe:2004rs}
\begin{equation} 
\label{eq:Sigma}
\Sigma = \frac{s_h}{h} \left( h^{(1)},h^{(2)},h^{(3)},h^{(4)}, h  \frac{c_h}{s_h} \right) \, ,
 \qquad  s_h\equiv \sin h/f\ ,\ \ \ h^2 \equiv \sum_{a=1}^4(h^{(a)})^2 \, ,
\end{equation}
the effective lagrangian \eq{efflag3} can be written in a SO(5) invariant way: 
\be
{\cal L}_{\rm eff}=  \bar \Psi_L^i \pslash \Big[\delta^{ij}\Pi^{L}_0(p)
 +\Sigma^i \Sigma^j  \Pi_{1}^L(p)\Big]\Psi_L^j
 +(L\rightarrow R)+
\bar \Psi_L^i \big[\delta^{ij} M_0(p)+M_1(p)\Sigma^i \Sigma^j\big] \Psi_R^j+h.c.\, ,
\label{ttt}
\ee
where here $i,j=1,...,5$ label SO(5) indices.
By projecting into the  SO(4)-preserving vacuum, $\langle \Sigma\rangle=(0,0,0,0,1)$, we can find the relations between the 
correlators of \eq{efflag3} and \eq{ttt}:
\be
\Pi_0^{L,R}=\Pi^{L,R}_{Q_4}\ ,\ \  \ \ \Pi_1^{L,R}=\Pi^{L,R}_{Q_1}-\Pi^{L,R}_{Q_4}\ ,\ \  \ \ M_0=M_{Q_4}\ ,\  \ \  \
M_1=M_{Q_1}-M_{Q_4}\, .
\label{relation}
\ee
The couplings of the SM top-quark,  $q_L=(t_L,b_L)$ and $t_R$, to the strong sector are defined by
their embedding into the external SO(5)-multiplets $\Psi_{L,R}$
since, according to  \eq{mixingsms},  this tells us to which operators they couple  to.
For the case $\Psi_{L,R}\in \bf 5$,  this is uniquely   given by
\be
\Psi_L=\frac{1}{\sqrt{2}}(b_L,-i b_L,t_L,it_L,0)\ ,\ \ \ 
\Psi_R=(0,0,0,0,t_R)\, .
\label{embeddings}
\ee
Plugging \eq{embeddings} into \eq{ttt},  expanding around the vacuum $\Sigma=(0,0,0,s_h,c_h)$ (that can be achieved from \eq{eq:Sigma}  after a proper  SU(2) rotation)
 and using \eq{relation}, one 
 obtains   \eq{efflag} and  \eq{selff}.
 
 \section*{Appendix B}

\begin{table}[h!]
\begin{center}
\begin{tabular}{|c||c|c|c|c|}
\hline
  $\bf r_L\diagdown\, {\bf r_R}$  &                  $\bf 1$ &  $\bf 5$         &        $\bf 10$       & $\bf 14$        \\
\hline
\hline
$\bf 5$       & $m=n=0$&$m=0,\, n=1$   &   $m=n=0$  &       $m=n=0$               \\
\hline
$\bf 10$         &$-$      &  $m=n=0$  &$\begin{array}{c}
{\it (i)}\ m=0,\,n=1\\
{\it (ii)}\ m=n=0
\end{array}$
             &                 $m=0,\, n=1$    \\
\hline
$\bf 14$         &$m=0,\, n=1$      &  $\begin{array}{c}
{\it (i)}\ m=n=0\\
{\it (ii)}\ m=0,\,n=2
\end{array}$  &       $m=0,\, n=1$                       &           $\begin{array}{c}
{\it (i)}\ m=0,\, n=1\\
{\it (ii)}\ m=1,\,n=1
\end{array}$    \\
\hline
\end{tabular}
\end{center}
\caption{Values of $m,n$   in  \eq{LR}  for different embeddings.}   
\label{default}
\end{table}%

In this appendix, we list the analog of the effective lagrangian \eq{efflag} for different embeddings of  $t_L$ and $t_R$ in SO(5) representations, $\bf r_L$ and $\bf r_R$ respectively (\eq{efflag} corresponds to $\bf r_{L,R}=5$). 

We split the lagrangian in  three parts, ${\cal L}_{\rm eff}={\cal L}^{\rm LL}_{\rm eff}+{\cal L}^{\rm RR}_{\rm eff}+{\cal L}^{\rm LR}_{\rm eff}$. For the $LL$ and $RR$ part, we have for the $\bf 10 =6\oplus4$ (under SO(4)):
 \bea
{\bf r_L=10:} \ \ \ \
{\cal L}_{\rm eff}^{\rm LL}&=&
   \bar b_L \pslash 
   \left(
   \Pi^{b_L}_0+ \frac{1}{2}c_h^{2}\, \Pi^{b_L}_{2}(p) 
    \right) b_L+
  \bar t_L \pslash
  \left( 
   \Pi^{t_L}_0+\left(\frac{1}{2}c_h^{2}- \frac{1}{4}s_h^{2}\right)\, \Pi^{t_L}_{2}(p) 
 \right)
t_L\, ,\nonumber\\
{\bf r_R=10:} \ \ \ \
{\cal L}_{\rm eff}^{\rm RR}&=&
  \bar t_R \pslash 
  \left(
   \Pi^{t_R}_0-2c_h\, \Pi^{t_R}_{1}(p) +\frac{1}{4}s_h^{2}\, \Pi^{t_R}_{2}(p) 
  \right)
  t_R\, ,
  \eea
   where
 \be
\Pi_0^{t_L,t_R}=1+\Pi^{L,R}_{Q_4}\  ,\ \  \ \ \Pi_1^{t_R}=-\frac{1}{2}\tilde{\Pi}^{R}_{Q_6}\  ,\ \  \ \ \Pi_2^{t_L,t_R}=2\left(\Pi^{L,R}_{Q_6}-\Pi^{L,R}_{Q_4}\right)\, ,
\label{relation10}
\ee
and $\Pi_{0,2}^{b_L}=\Pi_{0,2}^{t_L}$. We have included a term contracted with the Levi-Civita tensor that  in the corresponding of  \eq{efflag3} reads, $ \bar Q_{R\, ij}^{(6)} \pslash\,  \tilde{ \Pi}^R_{Q_6}(p) Q_{R\, kl}^{(6)}\epsilon ^{ijkl}$. For the ${\bf 14=9\oplus4\oplus1}$, we have
  \bea
{\bf r_L=14:} \ \ 
{\cal L}_{\rm eff}^{\rm LL}&=&
   \bar b_L \pslash 
   \left(
      \Pi^{b_L}_0+\frac{1}{2} c_h^{2}\, \Pi^{b_L}_{2}(p) 
   \right)
 b_L+
  \bar t_L \pslash
  \left( 
   \Pi^{t_L}_0 +\left(\frac{1}{2}c_h^{2}- \frac{1}{4}s_h^{2}\right)\, \Pi^{t_L}_{2}(p) + s_h^{2}c_h^{2}\, \Pi^{t_L}_{4}(p)\right)
t_L\, ,\nonumber\\
{\bf r_R=14:} \ \ 
{\cal L}_{\rm eff}^{\rm RR}&=&
  \bar t_R \pslash 
  \left( 
   \Pi^{t_R}_0 +\left(\frac{4}{5}c_h^{2}+\frac{1}{20}s_h^{2}\right)\, \Pi^{t_R}_{2}(p) +\frac{1}{20}\left(4c_h^{2}-s_h^{2}\right)^2\, \Pi^{t_R}_{4}(p)\right)  t_R\, ,
   \label{efflag30}
\eea
 where now
 \be
\Pi_0^{t_L,t_R}=1+\Pi^{L,R}_{Q_9}\  ,\ \  \ \ \Pi_2^{t_L,t_R}=2\left(\Pi^{L,R}_{Q_4}-\Pi^{L,R}_{Q_9}\right)\  ,\ \  \ \ \Pi_4^{t_L,t_R}=\frac{5}{4}\Pi^{L,R}_{Q_1}-2\Pi^{L,R}_{Q_4}+\frac{3}{4}\Pi^{L,R}_{Q_9}\, ,
\label{relation14}
\ee
and, as before, $\Pi_{0,2}^{b_L}=\Pi_{0,2}^{t_L}$. For the  $LR$ terms we have
\be
{\cal L}_{\rm eff}^{\rm LR}=
  s_h^{1+2m} c_h^{n}\  \bar t_L  M^t_{1}(p)\,  t_R
+h.c.\, ,
\label{LR}
\ee
where the values of $m$ and $n$ are given in Table~1.
In cases where  two terms    with different values of $m,n$ are possible, one of the two must  be suppressed  to  avoid  large flavor-violations;
this can be achieved by  imposing  extra parities to the strong sector, as advocated in Ref.~\cite{Mrazek:2011iu}.


\end{document}